# MediChain™: A Secure Decentralized Medical Data Asset Management System

Sara Rouhani
*Department of Computer Science*
*University of Saskatchewan*
Sasaktoon, Canada
sara.rouhani@usask.ca

Luke Butterworth
*Trioova*
Edmonton, Canada
luke@trioova.ca

Adam D. Simmons
*Trioova*
Edmonton, Canada
adam@trioova.ca

Darryl G. Humphery
*Trioova*
Edmonton, Canada
darryl@trioova.ca

Ralph Deters
*Department of Computer Science*
*University of Saskatchewan*
Sasaktoon, Canada
deters@cs.usask.ca

*Abstract*— The set of distributed ledger architectures known as blockchain is best known for cryptocurrency applications such as Bitcoin and Ethereum. These permissionless block chains are showing the potential to be disruptive to the financial services industry. Their broader adoption is likely to be limited by the maximum block size, the cost of the Proof of Work consensus mechanism, and the increasing size of any given chain overwhelming most of the participating nodes. These factors have led to many cryptocurrency blockchains to become centralized in the nodes with enough computing power and storage to be a dominant miner and validator. Permissioned chains operate in trusted environments and can, therefore, avoid the computationally expensive consensus mechanisms. Permissioned chains are still susceptible to asset storage demands and non-standard user interfaces that will impede their adoption. This paper describes an approach to addressing these limitations: permissioned blockchain that uses off-chain storage of the data assets and this is accessed through a standard browser and mobile app. The implementation in the Hyperledger framework is described as is an example use of patient-centered health data management.

## I. Introduction

The inability to access medical data (generated by a medical practitioner or device) in a timely and efficient manner is a long-standing problem in health care delivery globally. The electronic medical or health record (EHR and EMR) systems that have been put in place assume that a patient sees only the practitioners in one clinic (EMR) or in one political jurisdiction (EHRs managed by a province or state). These systems are practitioner-centric and the incentive to be interoperable is low. In reality, patients do not restrict themselves to a single doctor or clinic. The average U.S. patient has approximately 19 distinct medical records and in a 2010 survey reported seeing 18.7 different doctors during their lives [1]. This study did not take into account the broader health practitioner population such as pharmacists, physiotherapists, chiropractors, etc. Patients also don't restrict themselves to a single EHR coverage zone. People travel for leisure, for work, and they relocate for extended periods. People are also increasingly interested in managing their health with the help of data generated by wearable devices. This patient generated health data (PGHD) is easily shared with the device/service provider but not with health practitioners. Despite enormous investments by practitioners, medical facilities, and governments medical services delivery and health management continue to be hampered by inaccessibility to information, poor interoperability, inconsistent security of data assets, cumbersome privacy controls, and excluding patients from controlling access to data assets that they are the rightful owners of.

MediChain™ is a Hyperledger based blockchain application that addresses these issues with an architecture that allows MediChain™ to be both extensible and scalable. There are three main components to the system: a blockchain based access control module, off-chain data storage, and a patient-centered mobile and web user interface. In order to maintain a high level of application performance that is also economically viable, all of the data assets such as diagnostic images, lab test results, prescriptions, treatment plans, etc. are encrypted and stored on a secure cloud-based repository and a hash of the asset's URI is stored in MediChain™. This patent-pending method securely links on-chain and off-chain data without incurring the prohibitive computational and storage loads inherent in most blockchain architectures. With the use of smart contracts, the data asset owner (the patient) or their designate (e.g. a caregiver) can manage access permissions to each of their data assets in a flexible and secure manner. The web and mobile interfaces make the application easily accessible to patients, caregivers, and health practitioners.

In the following sections we first provide background information on the permissionless blockchain architectures Bitcoin and Ethereum and of the permissioned blockchain architecture Hyperledger (Section II). Section III provides a brief review of implementations similar to MediChain. The details of our implementation and the architectural model are given in Section IV. Section V provides a summary of our paper and discusses areas for future work.

## II. BACKGROUND

### A. Blockchain

The blockchain architecture pioneered by Bitcoin [2] is a special type of a distributed database which includes chronological blocks of transactions where each block includes a hash value of recently confirmed transactions and are linked to the previous block by adding the hash value of the previous block. The entire blockchain is stored on each of the participating nodes. Blockchain uses consensuses algorithms to verify transactions. Once a block is added to the blockchain it can never be changed or deleted without becoming out of synch with the copies on the distributed nodes. This provides three key properties of immutability, durability, and reliability. See Fig. 1 for a representation of a generic blockchain based application.

The blockchain architecture participants are not vetted prior to transacting and transactions are carried out without third party involvement. This is a permissionless environment where participants may be anonymous. These properties have fueled the use of blockchain technology for the creation and trading of cryptocurrencies.

In the 10 years since Bitcoin's emergence it has become the dominant cryptocurrency. Its popularity has exposed functional weaknesses. The compute cycles required for the Proof of Work consensus mechanism is expensive to execute, sometimes surpassing the value of the bitcoin transaction itself. The chain itself has grown to approximately 160 GB in size [3] and continues to increase. Updating a blockchain of this size in a timely manner requires considerable compute and storage capacity, stressing existing nodes. The block size allowed by Bitcoin is currently 1 MB and not likely to be increased in the near future. This restricts the use cases that the architecture can be applied to. It also limits the sophistication of the transaction that can be carried out.

A newer blockchain technology, Ethereum, does not have a maximum block size in MB. It is limited by the maximum cost of a single transaction [4]. Theoretically, market forces will keep an Ethereum-based chain functional and economical to operate. It is too early to tell if this will pan out given the amount of speculative buying of the ICO's associated with Ethereum blockchains. Ethereum has introduced functionality that allows for relatively limitless complexity in transaction rules: smart contracts. Smart contracts [5] create programmable triggers in the blockchain which automatically attempt to validate conditions and that execute the action(s) associated with the outcome. The structure of each type of smart contract is determined at the time the blockchain is deployed. Participants provide their own values for each term of the contract, providing for both flexibility and complexity in the transactions.

The Ethereum platform is gaining popularity for cryptocurrency related use cases. Its efforts to move away from the Proof of Work consensus mechanism may provide a means for reducing the computational load required to participate in an Ethereum chain. Even with Ethereum's advancements the limitations of permissionless blockchains is a hinderance for their use in a broad set of enterprise use cases.

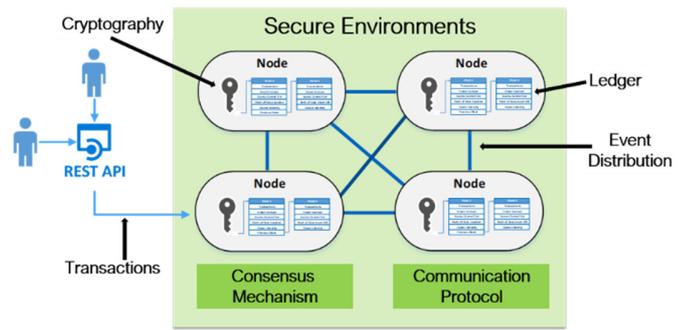

Fig. 1. Generic represention of Blockchain Application

The Hyperledger framework [8] managed by the Linux Foundation provides a promising alternative.

### B. Hyperledger Fabric

Hyperledger Fabric [6,7,8] is a blockchain platform that is permissioned; all nodes in the network have an identity. A Membership Services Provider (MSP) uses a public key infrastructure to issue each participant a cryptographic certificate. By adding cryptographic registration, identity management, and auditability a Hyperledger blockchain is able to function in a trusted environment. This provides the opportunity to utilize a consensus mechanism that is much lighter computationally than the Proof of Work mechanism. Hyperledger Fabric also adds the ability to create trusted subnetworks, called channels, that can establish shared ledgers with a defined set of nodes and transact to the exclusion of the rest of the blockchain. Fabric has smart contract functionality, Chaincode, which enables participants to execute complex transactions as in Ethereum. Fig. 2 shows a representative Hyperledger Fabric based application.

The Hyperledger Fabric architecture is modular allowing for flexibility in consensus mechanisms, membership service providers, certificate authorities, and interface SDK. Scalability is assisted by separating the Chaincode execution from the transaction ordering. These features make it possible to implement an efficient, secure and flexible blockchain platform that will scale with enterprise and global usage.

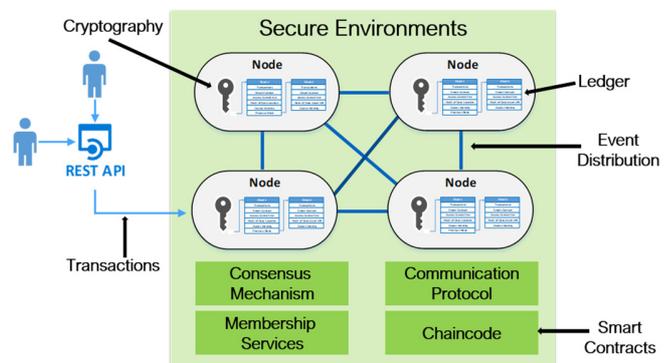

Fig. 2. Hyperledger based application

## C. Medical Access Control by Patient

A Hyperledger based blockchain seems the ideal means of solving the medical and health data asset challenges described in the Introduction. A permissioned blockchain with Chaincode smart contracts put the patient and caregiver in control, cryptography and immutability provide for privacy and security, and the permissioned architecture allows for trusted exchanges (transactions) between known participants. Interoperability challenges are bypassed as integration with EMRs and EHRs are not required. A browser and mobile app interface can provide simple drag-and-drop transfer of digital assets.

Hyperledger blockchains still have the challenge of the computational and storage cost of large block sizes. In many proposed enterprise use cases the size of the digital asset, and thereby the size of the block, that the blockchain is to manage would pose performance issues that would degrade user experience and drive up operational costs. Our implementation approach is designed to address this challenge.

## III. RELATED WORKS

The Personally Control Health Record (PCHR) [9] introduces a patient centric healthcare data sharing model based on both role-based and discretionary access control methods. In the PCHR model, patients have the ultimate authority in determining who can access their data. [14] Enigma is a decentralized computation platform which considers privacy along with computation and data storage to achieve privacy and scalability. Data is split into unmeaning chunks and each node has access to one chunk of data, unlike blockchain the data are not replicated and computed by every node. However, an external blockchain is employed to control the system, manage access control, and as a tamper-proof database of events. [11] This model uses a blockchain remove third party access to personal data. As a result, users are able to control access to their own data. The system has been implemented based on the combination of off-chain storage and storing a pointer to encrypted data on the Bitcoin blockchain. The blockchain portion of the system handles sharing and querying the data. Medrec [10] is another implementation of an access control system which has uses blockchain technology. Medrec is implemented using Ethereum technology with some modification of the mining process. It presents a reward-based mining method to motivate medical stakeholders to participate in the system and verify transactions as miners. [15] proposed a blockchain framework for sharing electronic medical data stored in the cloud repositories. The system is based on a permissioned blockchain therefore only authorized users would have access to the system by verification of their cryptographic keys. The performance testing based on comparison of this blockchain and bitcoin blockchain demonstrates a light and scalable design. [12] presents an implementation of role-based access control using the smart contracts and the challenge response protocol based on the Ethereum platform. The challenge-response protocol is designed to authenticate the ownership of roles and for user verification. RBAC-SC have focused on trans-organization access control, a user accesses a service of one organization based on his or her role in another organization. [13] utilizes blockchain for validating EHR records by using an attribute-based signature (ABS) method with multiple authorities. The evaluation they report illustrates that this system is robust against collusion attacks as well as for preserving privacy.

Although there are similar works in the scope of access control based on blockchain, MediChain™ is the first implementation of a functioning, real-world application based on Hyperledger Composer and Hyperledger Fabric to utilize a permissioned blockchain, to address privacy and efficiency issues in healthcare systems.

## IV. SYSTEM IMPLEMENTATION

Our system includes three main components: a permissioned blockchain, off-chain storage, and a patient-centric user experience accessed through a browser and mobile app interface. Fig. 3 provides an overview of our MediChain™ architecture as instantiated in the Trioova™ application. Each main component is described in sub-sections below.

### A. Permissioned Blockchain

Hyperledger Composer [17] was used to create the Business Network Archive (BNA) that determines the characteristics and capabilities of MediChain™. Composer is also used to deploy a runtime version of the BNA onto the Fabric instance. There are three main files created with Composer: a Model file, a Script file, and the Access Control List.

*1) Model File.*

The model file defines the Participants, Assets, and Transactions supported by MediChain™

*a) Participants:*

As a permissioned blockchain all users are authenticated by a third party (not shown in Fig. 3) before being given access to the chain. When a user is authenticated by the third party they are assigned a blockchain ID card by the Membership Services Provider that has been deployed on our Fabric instance. There are three participants types in our system: patient, caregiver, and health practitioner. An individual can register as one, two, or all three and unique IDs are assigned to each registration. Each participant type can be a data owner and a data requester. Data requesters are authenticated users who can submit transactions that request access to data assets of specific data owners.

*b) Assets:*

The assets in the Trioova™ application are medical and health data. Initially, each asset is considered a private asset. The owner is the only participant who has access to the asset. The owner of the data or their designated caregiver has the right to share a data asset with others via transactions.

*c) Transactions:*

Any request for a change to asset access starts with a transaction initiated by a data requester. The requests are managed in a publish-subscribe fashion.

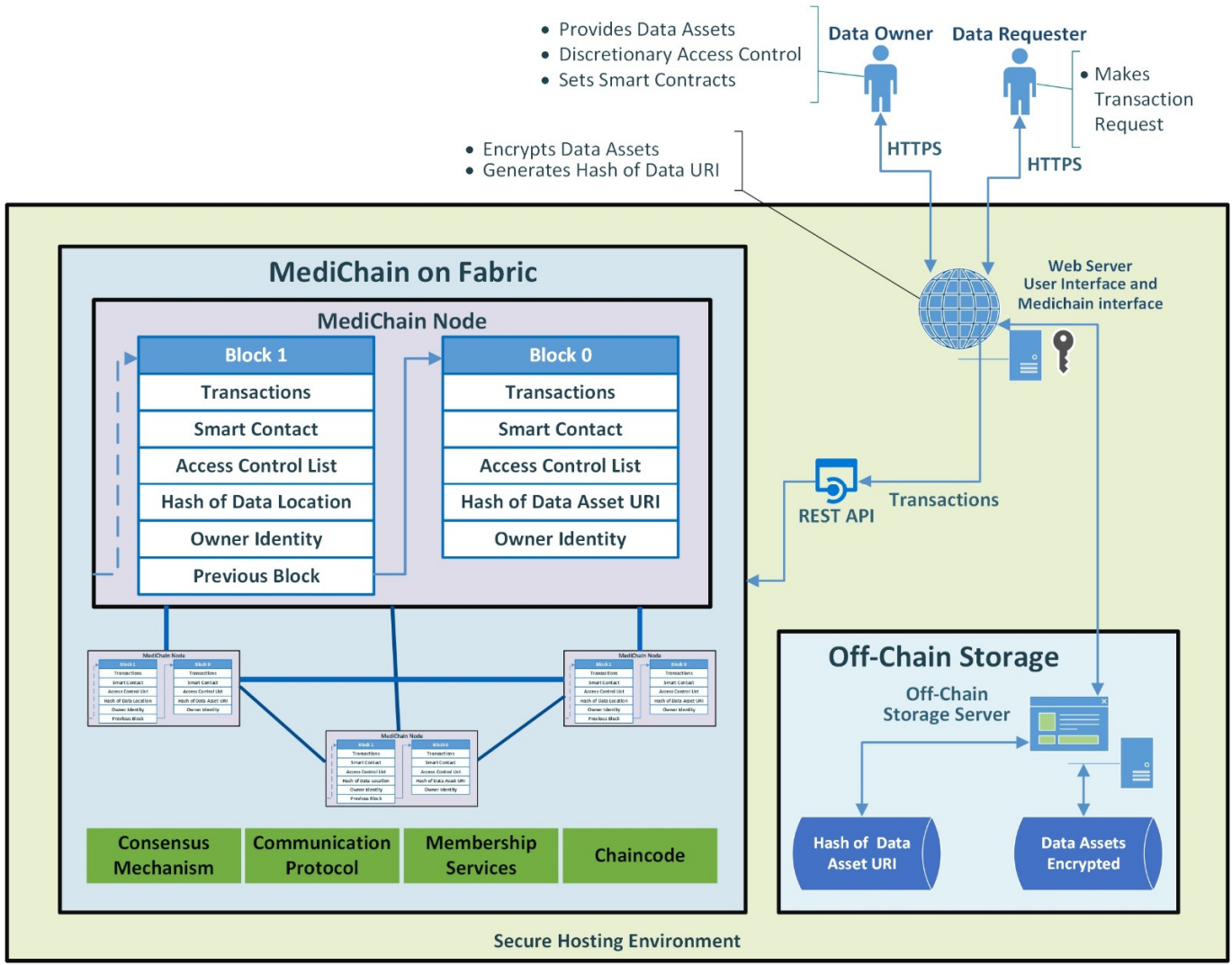

Fig. 3. MediChain[TM] architecture as instantiated in the Trioova[TM] application

*2) Script File (Chaincode Smart Contracts)*

Chaincode provides the smart contract functionality as described in the Ethereum discussion above. In the script file the specific conditions, or clauses, of each smart contract type are defined. The mobile or web interface enables the user to specify values for conditions in the contract. For example, a smart contract could be designed to provide time-based access to a data asset. The data requester would provide the identity of the asset and the proposed length of time (or date, etc.) for accessing the asset. The data owner can agree to these terms or propose modifications. When the conditions are confirmed the transaction is executed. Fig. 4 illustrates an example of this workflow.

*3) Access Control List*

Hyperledger Composer provides complex condition language that can be used to specify default policies and actions for the runtime BNA. One of these is the access control policy. The three main access control policies are Discretionary Access Control (DAC), Mandatory Access control (MAC) and Role Based Access Control (RBAC) [16]. The DAC is owner-centric which means that for any resource there is a specific owner as an entity in the system. The owner has full discretionary authority over who else can access the resource. DAC provides simplicity, flexibility, however, publication of access rights in DAC is infinite and difficult to predict. Contrary to DAC, MAC enforces access to resources via predefined administrator procedures and users and the owner of the resources are not able to change the access control. RBAC provides the same access permissions for everyone associated with a given role. For our application, we focus on DAC as our main access control policy since we believe that the owner of the data should have discretionary authority over their own data.

Hyperledger Composer is also used to define the allowable events and queries. After all of the initial entities and transactions have been set up an efficient logging and querying method is used to track all of the submitted transactions. Any owner has the ability to track and monitor any request. This provides the auditing capability.

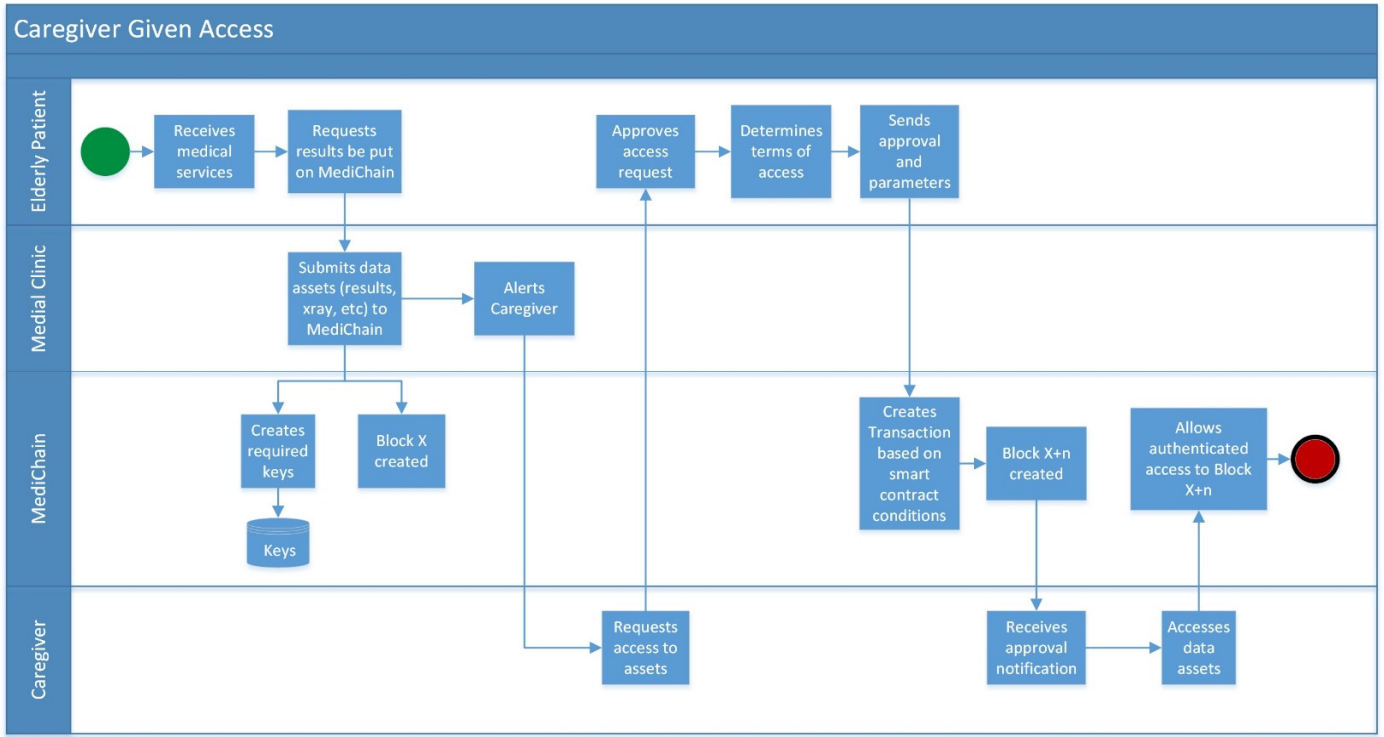

Fig. 4. Workflow with smart contract controled access.

*B. Off-chain Storage*

To maintain the performance and economic viability of a MediChain[TM] based application the medical and health data assets are stored in secure off-chain storage and the hash of the data asset's URI is the data element of the block that is committed to the chain. Transactions involving the data asset are signed with the private key of the asset owner (patient or caregiver). By using the hash of the URI and private key cryptography we minimize the size of each individual block and provide an additional layer of security for the data asset.

*C. Patient-centric User Experience*

Trioova[TM] is the web application and mobile app that is used to access the MediChain[TM] capabilities. The organizing principle is that the patient is always at the center of a circle of care. Patients choose who to include in their care circle and, at a granular level, what information is shared with each. For patients that need or want the assistance of a trusted party, the circle expands to include a caregiver that can act on the patient's behalf.

The TrioovaTM platform can also be used to find appropriate health practitioners, request appointments, and manage care plans across multiple medical and health practitioners. Fig. 5 shows the representative pages of the web and mobile apps.

The Trioova[TM] platform connects to the MediChain[TM] Fabric via a RESTful API. Hyperledger Composer provides a feature to call external APIs from transaction processor function or smart contracts. To enable modularity and allow for seamless extensibility, we developed a secure, robust, and scalable middleware component to connect users to the blockchain. This component is built, in part, with Hypertext Transfer Protocol Secure (HTTPS). With Trioova[TM], users can control access to their digital assets from any device that supports HTTPS requests.

V. SUMMARY AND FUTURE WORK

With MediChain[TM] patients can play a more active role in the health care, the management of their medical and health data, and the healthcare of their loved ones.

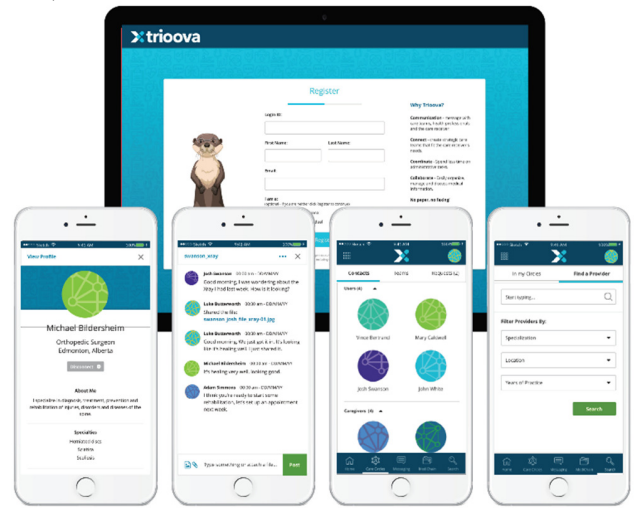

Fig. 5. Trioova[TM] mobile and web interfaces screen shot.

MediChain™ facilitates the efficient exchange of health data between patients, caregivers, and health practitioners while simultaneously enabling a secure protocol for the movement of private health data. This architecture is scalable at the enterprise and global level. It provides for superior ability to manage access to medical and health assets and enhanced security of those assets.

We believe that the current prototype can be generalized to any use case requiring the management of large digital assets such as music, visual arts, and records management.

Our future work will have two foci. One is performance testing of the MediChain™ based Trioova application when operating at scale. The second is measuring the application' and the blockchain to the most likely attack vectors. and is the next focus when operating at scale is the analysis of our system and exposing it to various security attacks and monitor the behavior of the system are imminent parts of our future work.

ACKNOWLEDGEMENTS

Development of MediChain™ and Trioova™ was supported in part by the Mitacs Accelerate program and Alberta Innovates